\documentclass[10pt,conference]{./sty/IEEEtran}

\usepackage{subfigure}
\usepackage{color}
\usepackage{dsfont}
\usepackage{cite}
\usepackage{graphicx}
\usepackage{subfigure}
\usepackage{url}
\usepackage{amsmath}
\usepackage{balance}

\usepackage{epsfig}
\usepackage{array}
\usepackage{psfrag}
\usepackage{amssymb}

\def\bx{{\bf x}}
\def\by{{\bf y}}
\def\bz{{\bf z}}

\def\bA{{\bf A}}
\def\bB{{\bf B}}
\def\bC{{\bf C}}

\def\bG{{\bf G}}
\def\bH{{\bf H}}
\def\bI{{\bf I}}
\def\bM{{\bf M}}
\def\bP{{\bf P}}
\def\bQ{{\bf Q}}

\def\bT{{\bf T}}
\def\bU{{\bf U}}

\def\bLambda{{\bf \Lambda}}

\def\bTheta{{\bf \Theta}}

\def\b0{{\bf 0}}

\def\bbI{{\mathcal{\boldsymbol I}}}

\newtheorem{theorem}{Theorem}
\newtheorem{lemma}{Lemma}

\begin{document}

\title{Asymptotic Capacity and Optimal Precoding Strategy of Multi-Level Precode \& Forward in Correlated Channels}

\author{
\authorblockN{Nadia Fawaz\authorrefmark{1},
Keyvan Zarifi\authorrefmark{2},
Merouane Debbah\authorrefmark{3} and
David Gesbert\authorrefmark{1}}
\authorblockA{\authorrefmark{1}Mobile Communications Department,
Eurecom Institute,
Sophia-Antipolis, France \\
\{nadia.fawaz,david.gesbert\}@eurecom.fr}
\authorblockA{\authorrefmark{2}INRS-EMT
\& Concordia University,
Montr\'{e}al, Canada \\
keyvan.zarifi@emt.inrs.ca}
\authorblockA{\authorrefmark{3}Alcatel-Lucent Chair on Flexible Radio,
SUPELEC,
Gif-sur-Yvette, France \\
merouane.debbah@supelec.fr}
}

\maketitle

\begin{abstract}
We analyze a multi-level MIMO relaying system where a multiple-antenna transmitter sends data to a multiple-antenna receiver through several relay levels, also equipped with multiple antennas. Assuming correlated fading in each hop, each relay receives a faded version of the signal transmitted by the previous level, performs precoding on the received signal and retransmits it to the next level.
Using free probability theory and assuming that the noise power at the relay levels - but not at the receiver - is negligible, a closed-form expression of the end-to-end asymptotic instantaneous mutual information is derived as the number of antennas in all levels grow large with the same rate. This asymptotic expression is shown to be independent from the channel
realizations, to only depend on the channel statistics and to also serve as the asymptotic value of the end-to-end average mutual information.
We also provide the optimal singular vectors of the precoding matrices that
maximize the asymptotic mutual information : the optimal transmit directions represented by the singular vectors of the precoding matrices are aligned on the eigenvectors of the channel correlation matrices, therefore they can be determined only using the known statistics of the channel matrices and do not depend on a particular channel realization.
\end{abstract}

\section{Introduction}\label{sec:Introduction}

Relay communication systems have recently attracted much attention
due to their potential to substantially improve the  signal
reception quality when the direct communication link between the
transmitter and the receiver is not reliable. Due to its major
practical importance as well as its significant technical challenge,
deriving the capacity - or bounds on the capacity - of various relay
communication schemes is growing to an entire field of research. Of
particular interest, is the capacity bounds for the systems in which
the transmission, reception, or the relay levels are equipped with
multiple antennas.
 Assuming fixed channel conditions, lower and upper bounds
on the capacity of multiple-input multiple output (MIMO) two-hop
relay channel have been derived in \cite{WZH}. Similar bounds have
also been obtained in the same paper on  the ergodic capacity when
the communication links undergo i.i.d. Rayleigh fadings. For a
two-hop relay system and the case when the source-relay and the
relay-destination channel matrices are perfectly known, the optimal
relay precoding matrix  is derived in \cite{TH07}. In \cite{BNOP06} the asymptotic capacity of MIMO two-hop
relay networks has been studied when the number of relay nodes grows to infinity while
the number of transmit and receive antennas remain constant. The
asymptotic capacity of the MIMO multi-hop amplify-and-forward relay
channels has been derived in \cite{YL07} when all channel links
experience i.i.d. Rayleigh fading while the number of transmit and
receive antennas as well as the number of relays at each hop go to
infinity with the same rate. The scaling behavior of the capacity of
MIMO two-hop relay channel has also been studied in \cite{VH07} for
the case when the source and destination jointly communicate to each
other through the relay layer. Other related problems have been
analyzed in, for instance, \cite{Mueller-2002},  \cite{MB07}, \cite{FT07}, \cite{Yang-Belfiore-2008}.

\begin{figure*}[tbp]
\centering
 \psfrag{x_0}{$\bx_0$}
\psfrag{x_1}{$\bx_1$}
\psfrag{x_2}{$\bx_2$}
\psfrag{x_(N-2)}{$\bx_{N-2}$}
\psfrag{x_(N-1)}{$\bx_{N-1}$}
\psfrag{y_0}{$\by_0$}
\psfrag{y_1}{$\by_1$}
\psfrag{y_2}{$\by_2$}
\psfrag{y_(N-2)}{$\by_{N-2}$}
\psfrag{y_(N-1)}{$\by_{N-1}$}
\psfrag{yN}{$\by_N$}
\psfrag{k_0}{$k_0$}
\psfrag{k_1}{$k_1$}
\psfrag{k_2}{$k_2$}
\psfrag{k_(N-2)}{$k_{N-2}$}
\psfrag{k_(N-1)}{$k_{N-1}$}
\psfrag{k_N}{$k_N$}
 \psfrag{x_0}{$\bx_0$}
\psfrag{H_1}{$\bH_1$}
\psfrag{H_2}{$\bH_2$}
\psfrag{H_(N-1)}{$\bH_{N-1}$}
\psfrag{H_N}{$\bH_N$}
\resizebox{16.5cm}{!}{\includegraphics{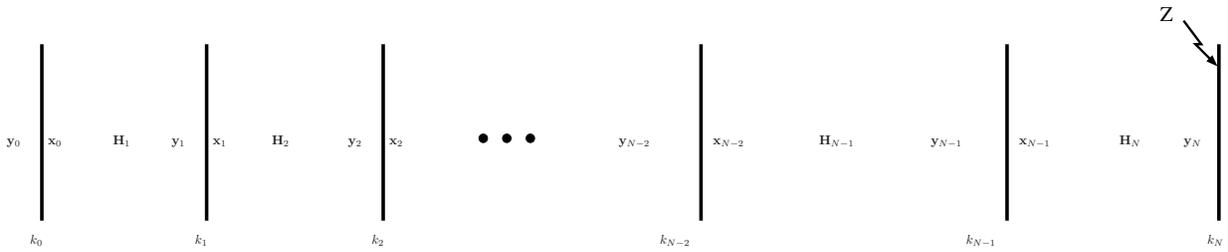}}
\caption{Multi-level Relaying System}
\label{figure1}
\end{figure*}

In this paper we study MIMO $N$-hop relay communication system
wherein data transmission from $k_0$ transmit antennas to $k_N$
receive antennas is made possible through $N-1$ relay levels each of
which are equipped with $k_i,~i=1,\ldots,N-1$ antennas. In this
transmission chain of $N+1$ levels it is assumed that the direct
communication link is only viable between two adjacent levels: Each
relay receives a faded version of the multi-dimensional signal
transmitted from the previous level and, after precoding, retransmits it
to the next level. We consider the case where all communication
links undergo Rayleigh flat fading and the fading channels in each
hop (in-between two adjacent levels) may be correlated while the fading
channels of any two different hops are uncorrelated.
Using the tools from free probability theory and assuming that the noise power at
the relay levels, but not at the receiver, is negligible, we derive
a closed-form expression for the end-to-end asymptotic instantaneous mutual
information between the transmitter and the receiver as the number
of antennas in all levels grows large with the same rate. This
asymptotic expression is shown to be independent from the channel
realizations and only depends on the channel statistics. Therefore,
as long as the statistical properties of the channel matrices at all
hops do not change, the instantaneous mutual information
asymptotically converges to the same deterministic expression for any
arbitrary channel realization. As a consequence, the so-obtained
expression also serves as the asymptotic value of the end-to-end average mutual
information between the transmitter and the receiver.
Then we obtain the optimal singular vectors of the precoding matrices that
maximize the asymptotic  mutual information. It is shown that, in the asymptotic regime, the optimal
singular vectors of the precoding matrices are also independent from
the channel realizations and can be determined only using the known
statistics of the channel matrices.

The rest of the paper is organized as follows. In section
\ref{sec:SysMod}, notations and the system model are
presented. In section \ref{sec:AsymptoticMutualInfo}, the end-to-end instantaneous mutual information in the asymptotic regime is derived, whereas the optimal singular vectors of the precoding matrices are given in section \ref{sec:OptimTxStrategy}. Numerical results are provided in section \ref{sec:NumRes} and lead to the concluding section \ref{sec:Conclusion}.

\section{System Model}\label{sec:SysMod}

Matrices and vectors are represented by boldface
uppercase. $\textbf{A}^T$, $\textbf{A}^\ast$, $\textbf{A}^H$ denote
the transpose, the conjugate and the transpose conjugate of matrix $\textbf{A}$. $tr(\textbf{A})$, $\det(\textbf{A})$ and $\|\textbf{A}\|_F =\sqrt{ tr(\textbf{A} \textbf{A}^H) }$ stand for trace, determinant and Frobenius norm of $\textbf{A}$.
$\textbf{I}_N$ is the identity matrix of size N.

Consider Fig.~\ref{figure1} that shows a multi-level relaying system with $k_0$ transmit antennas, $k_N$ receive antennas and $N-1$ relaying stages. The $i-$th  relay stage is equipped with $k_i$ antennas. We assume that noise power is zero at all relays while at the receiver we have
\begin{equation}
 {\rm E}\{\bz\bz^H\}=\sigma^2\bI=\frac{1}{\eta}\bI \label{noise}
\end{equation}
where $\bz$ is the circularly-symmetric zero-mean iid Gaussian receiver noise vector. 
This simplifying non-noisy-relays assumption is a first step toward more complete analysis where noisy relays will be considered in future work. (Remark: in \cite{Yang-Belfiore-2008} a multi-level AF relay network with iid Rayleigh fading is analyzed at high SNR and it is shown that at high SNR the colored noise at destination can be considered as white, which is equivalent to neglecting the noise at relays but not at the destination.)
Channel matrices are given by the Kronecker model:
\begin{equation}
 \bH_i=\bC_{r,i}^{1/2}\bTheta_i\bC_{t,i}^{1/2}\qquad i=1,\ldots,N\label{channels}
\end{equation}
where $\bC_{t,i}, \bC_{r,i}$ are the transmit and receive correlation matrices and $\bTheta_i$ are zero-mean iid Gaussian matrices, independent from each other. Moreover, denoting $[\bTheta_i]_{kl}$ the $(k,l)$ entry of $\bTheta_i$:
\begin{equation}
 {\rm E}\{|[\bTheta_i]_{kl}|^2\}=\frac{1}{k_i}\qquad i=1,\ldots,N\label{channelvars}
\end{equation}

The signal transmitted by the transmitter is $\bx_0 = \bP_0 \by_0$ where $\by_0$ is a circularly-symmetric zero-mean iid Gaussian signal vector such that
 ${\rm E}\{\by_0\by_0^H\}=\bI$.

Relay at level $i$ performs linear precoding on its received signal, by multiplying the received vector by a precoding matrix to form its transmitted signal. Therefore the vectors transmitted by the relays $\bx_i, i=1,\ldots, N-1$ are given by:
\begin{equation}
 \bx_i=\bP_i\by_i\qquad i=0,\ldots, N-1\label{xyrel}
\end{equation}
where $\by_i, i=1,\ldots, N-1$ are the signals received at each relaying level, and $\bP_i,$ are to-be-determined precoding matrices, respecting the per-node average power constraints:
\begin{equation}\label{eq:PowConstraints}
tr({\rm E}\{\bx_i \bx_i^H\}) \leq \mathcal{P}_i \qquad i=0,\ldots,N-1
\end{equation}
Amplify-and-Forward is a particular case where the precoding matrix is diagonal.

According to Fig.~\ref{figure1}, the signal received at the final receiver can be represented  by
\begin{eqnarray}\label{sigmodl}
 \by_N&=&\bH_N\bP_{N-1}\bH_{N-1}\bP_{N-2}\ldots\bH_2\bP_1\bH_1\bP_0\by_0+\bz\nonumber\\
&=&\bG_N\by_0+\bz
\end{eqnarray}
where
\begin{equation}\label{GN2}
\begin{split}
 \bG_N =&\bH_N\bP_{N-1}\bH_{N-1}\bP_{N-2}\ldots\bH_2\bP_1\bH_1\bP_0\\
 =&\bC_{r,N}^{1/2}\bTheta_N\bC_{t,N}^{1/2}\bP_{N-1}\bC_{r,N-1}^{1/2}\bTheta_{N-1}\bC_{t,N-1}^{1/2}\bP_{N-2}\ldots \\
 & \times \bC_{r,2}^{1/2}\bTheta_2\bC_{t,2}^{1/2}\bP_1\bC_{r,1}^{1/2}\bTheta_1\bC_{t,1}^{1/2}\bP_0
\end{split}
\end{equation}

We define the matrices :
\begin{eqnarray}
\bM_i&=&\bC_{t,i+1}^{1/2}\bP_i\bC_{r,i}^{1/2}\qquad i=1,\ldots,N-1\nonumber\\
\bM_0&=&\bC_{t,1}^{1/2}\bP_0\nonumber\\
\bM_N&=&\bC_{r,N}^{1/2}.\label{Ms}
\end{eqnarray}
and can then rewrite (\ref{GN2}) as:
\begin{equation}\label{GN4}
 \bG_N =\bM_N\bTheta_N\bM_{N-1}\bTheta_{N-1}\ldots\bM_2\bTheta_2\bM_1\bTheta_1\bM_0
\end{equation}

The dimensions of the matrices/vectors that are involved in our analysis are given below:

\begin{center}
\begin{tabular}{lll}
$\bx_i: k_i\times 1$ & $\by_i: k_i\times 1$ & $\bP_i: k_i\times k_i$\\
$\bH_i: k_i\times k_{i-1}$ & $\bC_{r,i}: k_i \times k_i$ & $\bC_{t,i}: k_{i-1} \times k_{i-1}$\\
$\bTheta_i: k_i\times k_{i-1}$ & $\bM_i: k_i \times k_i$ &
\end{tabular}
\end{center}

\section{Asymptotic Mutual Information}\label{sec:AsymptoticMutualInfo}

In this section, we derive a closed-form expression for the end-to-end asymptotic instantaneous mutual information between the transmitter and the receiver as the number of antennas in all levels grows large with the same rate.
In other words, we find the asymptotic instantaneous mutual information per dimension $\bbI$ \begin{equation}
 \bbI=\frac{1}{k_0}\log\det(\bI+\eta\bG_N\bG_N^H)
\end{equation}
when $k_0, k_1,\ldots, k_N$ go to infinity while
\begin{equation}
 \frac{k_i}{k_N}\rightarrow\rho_i\qquad i=0,\ldots, N
\end{equation}

The main result is summarized in the following theorem:

\smallskip

\begin{theorem}\label{th:Iasymptotic}
For the system described in section \ref{sec:SysMod}, as $k_0, k_1,\ldots, k_N$ go to infinity while $\frac{k_i}{k_N}\rightarrow\rho_i, i=0,\ldots, N$ the asymptotic end-to-end instantaneous mutual information per dimension $\bbI$ is given by
\begin{equation}\label{hh1}
\bbI=\frac{1}{\rho_0}\sum_{i=0}^N\rho_i{\rm E}\left\{\log\left(1+\frac{\eta}{\rho_{i+1}} h_i^N\Lambda_i\right)\right\}-N\frac{\log e}{\rho_0}\eta\prod_{i=0}^N h_i
\end{equation}
where $h_0,h_1,\ldots,h_N$ are the solutions of the system of $N+1$ equations
\begin{equation}\label{hh2}
\prod_{j=0}^N h_j=\rho_i{\rm E}\left\{\frac{h_i^N\Lambda_i}{\rho_{i+1}+\eta h_i^N\Lambda_i}\right\}\qquad i=0,\ldots,N
\end{equation}
and where  $\rm E$ is over $\Lambda_i$ with distribution  $F_{\bM_i^H\bM_i}(\lambda)$
\end{theorem}

\smallskip

The proof of this theorem uses tools from free probability theory. After introducing a few transformations and lemmas, we provide hereafter the main steps of the proof of \emph{Theorem \ref{th:Iasymptotic}}. For the full proof, the reader is referred to \cite{Fawaz-Zarifi-Debbah-2008}.

Before giving a sketch of the proof, we would like to point out that this expression of the asymptotic instantaneous mutual information is valid for any arbitrary set of matrices $\bP_i$.

Moreover this asymptotic expression depends only on the channel statistics and not on a particular realization of the channel. Thus when the
size of the system gets large, by knowing only the statistics of the
channel and not the instantaneous channel realization, it is still possible to optimize the instantaneous mutual information, making Theorem \ref{th:Iasymptotic} a powerful tool for optimizing the system performance. Actually, it is a powerful optimization tool even for a small number of antennas. Indeed as illustrated in section \ref{sec:NumRes}, experimental results show that the system behaves like in the asymptotic regime even for a small number of antennas. In other words, the asymptotic mutual information can be used to optimize the instantaneous mutual information of a finite-size system when transmitting nodes know only the statistics of the channel (the receiver is assumed to know the channel).

Finally, since the asymptotic mutual information depends only on channel statistics, as long as the statistical properties of the channel matrices do not vary at all hops, the instantaneous mutual information converges asymptotically to the same deterministic expression for any arbitrary channel realization. As a consequence, the asymptotic instantaneous end-to-end mutual information is also the asymptotic value of the average end-to-end  mutual information.

\subsection{Preliminaries}\label{sec:Preliminaries}

To prove \emph{Theorem \ref{th:Iasymptotic}}, we need to introduce the following transformations \cite{Mueller-2002}:
\begin{eqnarray}
 \Upsilon_{\bT}(s) & \triangleq & \int\frac{s\lambda}{1-s\lambda}dF_{\bT}(\lambda) \label{ups} \\
 S_{\bT}(z)& \triangleq & \frac{z+1}{z}\Upsilon_{\bT}^{-1}(z) \mbox{ , \emph{S-transform}} \label{Strans}
\end{eqnarray}
where $\bT$ is a Hermitian matrix, $F_{\bT}$ is the asymptotic eigenvalue distribution of $\bT$ as its dimensions go to infinity and $\Upsilon^{-1}(\Upsilon(s))=s$.

Finally, we need the following lemmas:

\smallskip

\begin{lemma}[{\cite[Eq.(15)]{Mueller-2002}}]\label{lem:Mueller}
  Given an $m \times n$ matrix A with $\frac{n}{m}=\xi$
    \begin{equation}\label{ssh}
    S_{\bA\bA^H}(z)=\frac{z+1}{z+\xi}S_{\bA^H\bA}\left(\frac{z}{\xi}\right)
    \end{equation}
\end{lemma}

\smallskip

\begin{lemma}[{\cite{Debbah-Hachem-Loubaton-2003}}]\label{lem:Debbah}
   Let $\bTheta$ be a zero-mean iid Gaussian matrix and $\bT_1$ and $\bT_2$ be Hermitian matrices independent from $\bTheta$ and with proper dimensions such that $\bTheta\bT_1\bTheta^H\bT_2$ is meaningful. Then, as the dimensions of the matrices go to infinity, $\bTheta\bT_1\bTheta^H$ and $\bT_2$ become asymptotically free, thus,
    \begin{equation}\label{freeness}
    S_{\bTheta\bT_1\bTheta^H\bT_2}(z)=S_{\bTheta\bT_1\bTheta^H}(z)S_{\bT_2}(z)
    \end{equation}
\end{lemma}

\smallskip

\begin{lemma}[{\cite[Theorem 1]{Mueller-2002}}]\label{lem:Mueller2}
    Consider an $m \times n$ matrix $\bB$ with zero-mean iid entries with variance $\frac{1}{n}$. Assume that the dimensions go to infinity while $\frac{m}{n}\rightarrow\zeta$, then
    \begin{equation}\label{sbazi}
    S_{\bB\bB^H}(z)=\frac{1}{1+\zeta z}
    \end{equation}
\end{lemma}

\subsection{Main Steps of proof of Theorem \ref{th:Iasymptotic}}\label{sec:Roadmap}

The proof of \emph{Theorem \ref{th:Iasymptotic}} goes through four steps as follows:

\begin{itemize}
\item \textbf{First step :} Obtain $S_{\bG_N\bG_N^H}(z)$

Using \emph{Lemmas \ref{lem:Mueller}, \ref{lem:Mueller2}, \ref{lem:Debbah}} we show the following theorem:

\medskip

\begin{theorem}\label{th:SGG}
 As $k_i,i=0,\ldots,N$ go to infinity with the same rate, the S-transform of $\bG_N\bG_N^H$ is given by:
\small{
\begin{equation}\label{theo1}
 S_{\bG_N\bG_N^H}(z)=S_{\bM_N^H\bM_N}(z)\prod_{i=1}^N\frac{\rho_i}{z+\rho_{i-1}} S_{\bM_{i-1}^H\bM_{i-1}}\left(\frac{z}{\rho_{i-1}}\right)
\end{equation}
}
\end{theorem}

\item \textbf{Second step :} Use $S_{\bG_N\bG_N^H}(z)$ to find $\Upsilon_{\bG_N\bG_N^H}(z)$

\medskip

\begin{theorem}\label{th:UpsilonGG}
 Let us define $\rho_{N+1}=1$. We have
\begin{equation}\label{theo2}
 s\Upsilon_{\bG_N\bG_N^H}^N(s)=\prod_{i=0}^N\rho_{i+1}\Upsilon^{-1}_{M_i^HM_i}\left(\frac{\Upsilon_{\bG_N\bG_N^H(s)}}{\rho_i}\right)
\end{equation}
\end{theorem}

\item \textbf{Third step :} Use $\Upsilon_{\bG_N\bG_N^H}(z)$ to obtain $d\bbI/d\eta$.

First we note that
\begin{eqnarray}\label{Iint}
\bbI&=&\frac{k_N}{k_0}\frac{1}{k_N}\sum_{i=1}^{k_N}\log(1+\eta\lambda_i(\bG_N\bG_N^H))\nonumber\\
&=&\frac{k_N}{k_0}\int\log(1+\eta\lambda)dF^{k_N}_{\bG_N\bG_N^H}(\lambda)\nonumber\\
&\stackrel{a.s.}{\rightarrow}&\frac{1}{\rho_0}\int\log(1+\eta\lambda)dF_{\bG_N\bG_N^H}(\lambda)
\end{eqnarray}
where $F^{k_N}_{\bG_N\bG_N^H}(\lambda)$ is the empirical (non-asymptotic) eigenvalue distribution of $\bG_N\bG_N^H$.
Due to (\ref{Iint}), in the asymptotic regime, the derivative of the mutual information with respect to $\eta$ is linked to $\Upsilon_{\bG_N\bG_N^H}(z)$:
\begin{eqnarray}\label{theo33}
 \frac{d\bbI}{d\eta}&=&\frac{1}{-\rho_0\eta \ln 2}\Upsilon_{\bG_N\bG_N^H}(-\eta).
\end{eqnarray}

Thus using Theorem \ref{th:UpsilonGG}, we show the following theorem:

\medskip

\begin{theorem}\label{th:dIdeta}
In the asymptotic regime, as $k_0, k_1,\ldots, k_N$ go to infinity while $\frac{k_i}{k_N}\rightarrow\rho_i, i=0,\ldots, N$, the derivative of the instantaneous mutual information is given by:
\begin{equation}\label{theo3}
\frac{d\bbI}{d\eta}=\frac{1}{\rho_0\ln2} \prod_{i=0}^N h_i
\end{equation}
where $h_0,h_1,\ldots,h_N$ are the solutions of the system of $N+1$ equations
\begin{equation}\label{theo32}
\prod_{j=0}^N h_j=\rho_i{\rm E}\left\{\frac{h_i^N\Lambda_i}{\rho_{i+1}+\eta h_i^N\Lambda_i}\right\}\qquad i=0,\ldots,N.
\end{equation}
The expectation is over $\Lambda_i$ with distribution $F_{\bM_i^H\bM_i}(\lambda)$.
\end{theorem}

\item \textbf{Fourth step :} Integrate $d\bbI/d\eta$ to get $\bbI$

Since primitive functions of $\frac{d\bbI}{d\eta}$ differ by a constant, the constant was chosen such that the mutual information (\ref{hh1}) is null when SNR $\eta$ goes to zero : $\lim_{\eta \rightarrow 0} \bbI(\eta) = 0$.

\end{itemize}

\section{Optimal Transmission Strategy at Source and Relays}\label{sec:OptimTxStrategy}

In previous section, the asymptotic mutual information (\ref{hh1}), (\ref{hh2}) was derived considering arbitrary precoding matrices $\bP_i, i\in\{0,\ldots,N-1\}$.
In this section, we analyze the optimal linear precoding strategies $\bP_i, i\in\{0,\ldots,N-1\}$ at source and relays that allow to maximize the mutual information.

We characterize the optimal transmit directions, meaning the singular vectors of the precoding matrices at source and relays, that maximize the asymptotic mutual information. It turns out that those transmit direction are also the ones that maximize the average mutual information of finite size systems, i.e. when $k_0, k_1,\ldots, k_N$ are finite.
In future work, using the results on the optimal directions of transmission and the asymptotic mutual information (\ref{hh1}), (\ref{hh2}), we intend to work out the optimal power allocation in the asymptotic regime.

The main result of this section is given by the theorem:

\smallskip

\begin{theorem}\label{th:txDirections}
For $i\in\{1,\ldots,N\}$ let $\bC_{t,i}=\bU_{t,i}\bLambda_{t,i}\bU_{t,i}^H$ and $\bC_{r,i}=\bU_{r,i}\bLambda_{r,i}\bU_{r,i}^H$ be the eigenvalue decompositions of the covariance matrices $\bC_{t,i}$ and $\bC_{r,i}$ , where $\bU_{t,i}$ and $\bU_{r,i}$ are unitary and $\bLambda_{t,i}$ and $\bLambda_{r,i}$ are diagonal, with their respective eigenvalues ordered in decreasing order. Then the optimal linear precoding matrices, that maximize the asymptotic mutual information under power constraints (\ref{eq:PowConstraints}) can be written
\begin{equation}
\begin{split}
\bP_0 & =\bU_{t,1}\bLambda_{P_0} \\
\bP_i & =\bU_{t,i+1}\bLambda_{P_i}\bU_{r,i}^H \mbox{ , for } i\in\{1,\ldots,N-1\}
\end{split}
\end{equation}
where $\bLambda_{P_i}$ are diagonal matrices with complex elements.
Moreover, those precoding matrices are also the ones which maximize the average mutual information of a finite size (non-asymptotic) system.
\end{theorem}

\smallskip

This theorem means that the transmit directions at source and relays maximizing the asymptotic mutual information are such that:
\begin{itemize}
  \item the source should align the transmit covariance matrix $\bQ =\bP_0 \bP_0^H$ on the eigenvectors of the transmit correlation matrix of the channel $\bH_1$ to the first relaying layer
  \item relay $i$ should align the precoding matrix $\bP_i$ on the eigenvectors $\bU_{r,i}$ of receive correlation matrix of channel $\bH_i$ on the right, and on the eigenvectors $\bU_{t,i+1}$ of the transmit correlation matrix of channel $\bH_{i+1}$ on the left.
  \item the problem of optimizing $\bP_i$ can be divided into two decoupled problems: optimizing the transmit directions on one hand, and optimizing the transmit powers on the other hand.
\end{itemize}
For a detailed proof of \emph{Theorem \ref{th:txDirections}}, we once again refer the readers to \cite{Fawaz-Zarifi-Debbah-2008}. Nevertheless, we would like to draw the attention of the reader on two points.

First, the proof of this theorem does not rely on the expression of the asymptotic mutual information given in (\ref{hh1}) and is independent of \emph{Theorem \ref{th:Iasymptotic}}. On the contrary, the theorem is first proved for the non-asymptotic regime, and the result is shown to still hold in the limit when the dimensions increase. Nevertheless by combining \emph{Theorem \ref{th:Iasymptotic}} and \emph{Theorem \ref{th:txDirections}}, the ultimate objective is to find the optimal precoding matrices using only knowledge of the statistics of the channel.

Second, as in the proofs in \cite{Jafar-Goldsmith-2004} for the average mutual information of multiple-antenna single-user case with covariance knowledge at transmitter, or in \cite{Soysal-Ulukus-2007} for the average mutual information in the multiple-antenna multi-user case also with covariance knowledge at transmitter, both without relaying, or in \cite{TH07} for the two-hop relay system with full CSI at the relay, our proof of Theorem \ref{th:txDirections} is based on Theorem H.1.h in \cite{Marshall-Olkin-1979}.

\section{Numerical Results}\label{sec:NumRes}

In this section, we present numerical results to validate \emph{Theorem \ref{th:Iasymptotic}} and to show that even for a small number of antenna, the behavior of the system is close to the behavior in the asymptotic regime, making \emph{Theorem \ref{th:Iasymptotic}} a useful tool for optimization of finite-size systems as well as large networks.

\begin{figure}[htbp]
  \centering
  \includegraphics[width=0.98\columnwidth]{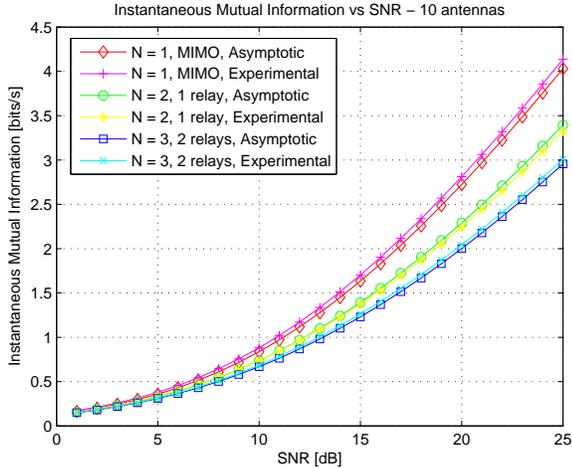}\\
  \caption[Instantaneous Mutual Information, 10 antennas]{Asymptotic Mutual Information and Instantaneous Mutual Information with K = 10 antennas, for MIMO, 1 relay level, and 2 relay levels}
  \label{fig:InstantMutInfo-K10}
\end{figure}

\begin{figure}[htbp]
  \centering
  \includegraphics[width=0.98\columnwidth]{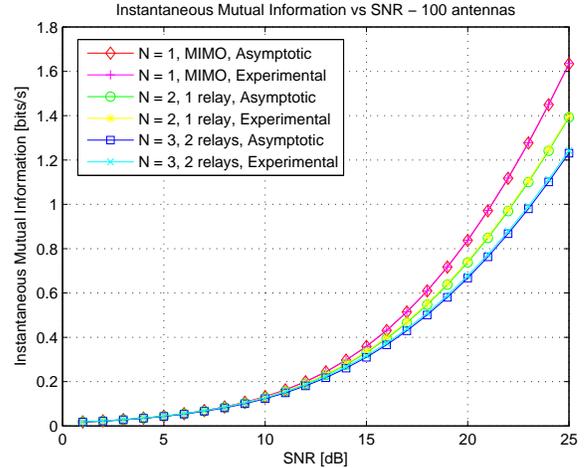}\\
  \caption[Instantaneous Mutual Information, 100 antennas]{Asymptotic Mutual Information and Instantaneous Mutual Information with K = 100 antennas, for MIMO, 1 relay level, and 2 relay levels}
  \label{fig:InstantMutInfo-K100}
\end{figure}

Fig. \ref{fig:InstantMutInfo-K10} plots the asymptotic mutual information from \emph{Theorem 1} as well as the instantaneous mutual information obtained for an arbitrary channel realization ('experimental' curves) for a system with 10 antennas at transmitter, receiver and each relay level, and 1,2 or 3 hops, the case $N=1$ hop corresponding to a MIMO channel. Fig. \ref{fig:InstantMutInfo-K100} plots the same type of curves, for a system with 100 antennas at each levels. Equal power allocation, i.e. matrices $P_i$ proportional to the identity matrix, as well as non-correlated channels, i.e. channel correlation matrices all equal to identity, were considered in these simulations, whose purpose is mainly to validate the formula in Theorem \ref{th:Iasymptotic}, not to optimize the system.
We would like to point out that plotting the experimental curves for different channel realizations gave similar results, and that for the sake of clarity and conciseness, we exhibit the experimental curves only for one realization.
Fig. \ref{fig:InstantMutInfo-K100} shows the perfect match between the instantaneous mutual information of arbitrary channel realization and the asymptotic mutual information, validating the formula for large dimensions of the network. On the other hand Fig. \ref{fig:InstantMutInfo-K10} shows that the instantaneous mutual information of a system with a small number of antennas behaves very closely to the asymptotic regime, justifying the usefulness of the asymptotic formula even for optimizing systems with small dimensions.

\section{Conclusion}\label{sec:Conclusion}
We studied a multi-level MIMO relay network, in correlated fading, where relays perform linear precoding on their received signal before retransmission.
On one hand, using free probability theory, we derived a closed-form expression of the end-to-end instantaneous mutual information in the asymptotic regime when the number of antennas at all levels goes to infinity with same rate. This expression turns out to depend only on channel statistics and not on particular channel realizations. We also showed that multi-level networks with finite dimensions behave closely to the asymptotic regime, even for a small number of antennas, making the asymptotic mutual information a powerful tool for optimizing the instantaneous mutual information of finite dimensions systems with only statistical knowledge of the channel.
On the other hand, we showed that the precoding matrices that maximize the asymptotic mutual information, have a particular form: the precoding matrices, through their singular vectors, must be aligned on the eigenvectors of the channel transmit and receive correlation matrices.
Combining asymptotic mutual information and optimal directions of transmissions, future work will focus on optimizing the power allocations, so as to find the precoding matrices optimizing the 
mutual information with only statistical channel knowledge.

\section*{Acknowledgment}

The authors would like to thank the French Defense Body DGA, BIONETS project (FP6-027748, www.bionets.eu) and Alcatel-Lucent within the Alcatel-Lucent Chair on flexible radio at SUPELEC for supporting this work.


\bibliographystyle{./Biblio/IEEEtran}

\bibliography{./Biblio/IEEEabrv,./Biblio/bibLargeNet}

\end{document}